\documentclass[12pt]{article}
\usepackage{a4wide}
\usepackage{amssymb}
\begin{document}
{\renewcommand{\thefootnote}{\fnsymbol{footnote}}
\hfill  CGPG--01/5--2\\
\medskip
\hfill gr--qc/0105113\\
\medskip
\begin{center}
{\LARGE  The Semiclassical Limit of Loop Quantum Cosmology}\\
\vspace{1.5em}
Martin Bojowald\footnote{e-mail address: {\tt bojowald@gravity.phys.psu.edu}}
\\
\vspace{0.5em}
Center for Gravitational Physics and Geometry,\\
The Pennsylvania State
University,\\
104 Davey Lab, University Park, PA 16802, USA\\
\vspace{1.5em}
\end{center}
}

\setcounter{footnote}{0}

\newcommand{\case}[2]{{\textstyle \frac{#1}{#2}}}
\newcommand{\lP}{l_{\mathrm P}}
\newcommand{\HE}{H^{({\rm E})}}

\newcommand{\md}{{\mathrm{d}}}
\newcommand{\tr}{\mathop{\mathrm{tr}}}
\newcommand{\sgn}{\mathop{\mathrm{sgn}}}

\newcommand*{\R}{{\mathbb R}}
\newcommand*{\N}{{\mathbb N}}
\newcommand*{\Z}{{\mathbb Z}}

\begin{abstract}
 The continuum and semiclassical limits of isotropic, spatially flat
 loop quantum cosmology are discussed, with an emphasis on the role
 played by the Barbero--Immirzi parameter $\gamma$ in controlling
 space-time discreteness. In this way, standard quantum cosmology is
 shown to be the simultaneous limit $\gamma\to0$, $j\to\infty$ of loop
 quantum cosmology. Here, $j$ is a label of the volume eigenvalues, and
 the simultaneous limit is technically the same as the classical limit
 $\hbar\to0$, $l\to\infty$ of angular momentum in quantum
 mechanics. Possible lessons for semiclassical states at the dynamical
 level in the full theory of quantum geometry are mentioned.
\end{abstract}

\section{Introduction}

Quantum geometry \cite{Nonpert,Rov:Loops}, one of the candidates for a
quantum theory of gravity, faces two main problems: the understanding
of its dynamics and of its classical limit. At present there are
candidates for a quantization of the Hamiltonian constraint
\cite{QSDI,Pullin}, which governs the dynamics, and proposals for the
construction of semiclassical states at the kinematical level
\cite{GCSI,CohState,StatGeo}. Both issues are complicated not only by
conceptual problems, but also by the fact that dealing with a general
state without any symmetry is technically difficult. In the first case
(understanding the dynamics), symmetric models \cite{SymmRed} of
homogeneous geometries \cite{cosmoI,IsoCosmo} have already proved to
be useful \cite{Sing,DynIn}. In this note, we will discuss the issue
of semiclassical states at the dynamical level.

A common concept of quantum mechanics in this context is the WKB
approximation. Intuitively, this gives an approximate solution to the
evolution equation of a wave function in a regime where the wave
propagation can be described by the motion of a particle (analogous to
the ray approximation in optics). Its basic condition is that there is
a well-defined, locally almost constant wave length in a neighborhood
of any point (which lies in superspace in the case of
gravity). Already here, we can see that there are necessary
modifications in the case of quantum geometry: since superspace is now
discrete \cite{AreaVol,Vol2}, the wavelength cannot be smaller than a
certain scale proportional to the Planck length $\lP$. At this point
the Barbero--Immirzi parameter $\gamma$ enters because it determines
this scale as $\sqrt{\gamma}\lP$, which can, e.g., be seen in the
isotropic volume spectrum \cite{cosmoII}
\begin{equation}\label{Vj}
 V_j=(\gamma\lP)^{\frac{3}{2} }\sqrt{\case{1}{27}
 j(j+\case{1}{2})(j+1)}\quad,\quad j\in\case{1}{2}\N_0\,.
\end{equation}
Similarly, $\gamma$ enters the spectra of other geometric operators
and thus controls the discreteness of geometry. Its physical magnitude
can be fixed from calculations of black hole entropy
\cite{ABCK:LoopEntro,IHEntro} which yields a value of the order
one. Note that, contrary to widespread statements in the literature,
the existence of a length scale $\lP=\sqrt{\kappa\hbar}$ does not by
itself guarantee space-time discreteness, as evidenced by standard
quantum cosmology \cite{DeWitt,Misner} which has continuous space
(scale factor) and time. (The same remark applies to quantum
electrodynamics in which the existence of a fundamental charge
$q=\sqrt{\alpha\hbar}$ does not by itself imply that the charge
spectrum must be discrete.) Only a non-zero value of $\gamma$ in
quantum geometry and loop quantum cosmology leads to a discrete volume
spectrum and discrete time evolution \cite{cosmoIV}. In fact, we will
show here in the isotropic, spatially flat case that standard quantum
cosmology is the $\gamma\to0$ (combined with $j\to\infty$) limit of
loop quantum cosmology (just as, e.g., Newtonian mechanics is the
$c^{-1}\to0$ limit of special relativity). In this way, $\gamma$
acquires the role of a useful and potentially fundamental parameter:
Classical equations of motion can be derived in a combined continuum
and semiclassical limit; in addition to ordinary quantum corrections
which vanish for $\hbar\to0$, there will be new corrections due to the
underlying discreteness which vanish for $\gamma\to0$ even if $\hbar$
is fixed.

\section{Isotropic Canonical Quantum Cosmology}

As is usual in quantum cosmology, we will work in a metric (or triad)
representation for the wave function $\psi$ of a universe in standard
quantum cosmology. More precisely, since the fundamental metrical
object in quantum geometry is a densitized triad which in our case can
be written as $E_i^a=p\Lambda^I_i X^a_I$ with an internal
$SU(2)$-triad $\Lambda_i^I$ and left invariant (with respect to the
homogeneity group) vector fields $X_I^a$, a wave function will be
represented as a function $\psi(p,\phi)$ where $\phi$ denotes matter
fields. Standard quantum cosmology only makes sense in the regime of
positive $p$ (whereas in loop quantum cosmology it is possible to
evolve a state to negative $p$ without encountering a singularity
\cite{Sing}), where the relation to the scale factor $a$ and space
volume $V$ is given by $p=a^2=V^{\frac{2}{3}}$.

Because the spectrum of $\hat{p}$ is discrete in quantum geometry, the
analog of $\psi(p,\phi)$ in loop quantum cosmology is a discrete wave
function $s_n(\phi)$ defined for integers $n\in\Z$. Using the identity
$|n|=2j+1$ and (\ref{Vj}) one obtains the relation
\begin{equation}\label{n}
 n=6\gamma^{-1}\lP^{-2}p
\end{equation}
for large positive $n\gg1$.

As canonically conjugate momentum to $p$ we have the connection
coefficient $c$ (from $A_a^i=c\Lambda_I^i\omega^I_a$ with left
invariant one-forms $\omega^I_a$ dual to $X_I^a$) fulfilling
$\{c,p\}=\frac{1}{3}\gamma\kappa$ ($\kappa=8\pi G$ is the
gravitational constant). For spatially flat cosmological models $c$ is
proportional to the extrinsic curvature of a spatial slice. In these
variables, the Hamiltonian constraint is given by
\begin{equation}
 H=-6\kappa^{-1}\gamma^{-2}c^2\sqrt{p}\,,
\end{equation}
which is obtained\footnote{Recall that in the full theory the
Lorentzian Hamiltonian constraint, multiplied by the determinant of
the co-triad, can be decomposed into a term which is quadratic in the
connection and triad and a non-polynomial term (see formulae
(\ref{Eucl}), (\ref{Ext})).} by adding the negative of the ``Euclidean
part'' $-\HE=6\kappa^{-1}c^2\sqrt{p}$ and the ``extrinsic curvature
part'' $P= -6(1+\gamma^{-2}) \kappa^{-1}c^2\sqrt{p}$.

Note that $H$ is proportional to $\gamma^{-2}$ thanks to the fact that
the $\gamma$-independent parts in $\HE$ and $P$ cancel. This leads to
a $\gamma$-independent Hamiltonian constraint equation in standard
quantum cosmology, which is obtained by quantizing
$\hat{c}=-\frac{1}{3}i\gamma\lP^2\md/\md p$,
\begin{equation}\label{HWdW}
 \kappa\hat{H}_{\rm WdW}\psi(p,\phi)\equiv
 \case{2}{3}\lP^4\frac{\md^2}{\md p^2}
 (\sqrt{p}\psi(p,\phi))= -\kappa\hat{H}_{\phi}\psi(p,\phi)
\end{equation}
using an arbitrary matter Hamiltonian $\hat{H}_{\phi}$ (which also
depends on $p$). In general, however, the constraint operator will be
$\gamma$-dependent when the conditions of spatial flatness or
homogeneity are dropped; we will comment on the implications later.

In loop quantum cosmology, which is closer to the full theory of
quantum cosmology, the constraint equation looks more complicated. It
is directly derived from an adaptation of Thiemann's operator
\cite{QSDI} to isotropy \cite{cosmoIII} and takes the form of a
difference equation for $s_n(\phi)$ (for details we refer to
\cite{IsoCosmo}):
\begin{eqnarray}
 \kappa(\hat{H}s)_n(\phi) &=& 3\gamma^{-1}\lP^{-2}
  \left(\case{1}{4}(1+\gamma^{-2})
  \left(V_{\frac{1}{2}(n+8)}- V_{\frac{1}{2}(n+8)-1}\right) k_{n+8}^+
  k_{n+4}^+ s_{n+8}(\phi)\right.\nonumber\\
 &&- \left(V_{\frac{1}{2}(n+4)}-
  V_{\frac{1}{2}(n+4)-1}\right) s_{n+4}(\phi)\nonumber\\
 && -2\left(V_{\frac{1}{2}n}- V_{\frac{1}{2}n-1}\right)
  \left(\case{1}{8} (1+\gamma^{-2})(k_n^-k_{n+4}^++
  k_n^+k_{n-4}^-)-1\right) s_n(\phi)\nonumber\\
 &&- \left(V_{\frac{1}{2}(n-4)}-
  V_{\frac{1}{2}(n-4)-1}\right) s_{n-4}(\phi)\nonumber\\
 && + \left.\case{1}{4}(1+\gamma^{-2})
  \left(V_{\frac{1}{2}(n+8)}- V_{\frac{1}{2}(n-8)-1}\right) k_{n-8}^-
  k_{n-4}^- s_{n-8}(\phi)\right)\nonumber\\
 &=& -\kappa\hat{H}_{\phi}(n) s_n(\phi)\,. \label{Hloop}
\end{eqnarray}
Here, the coefficients $k_n^{\pm}$, whose explicit form in terms of
the volume eigenvalues (\ref{Vj}) can be found in \cite{IsoCosmo},
come from the extrinsic curvature operator and rapidly approach the
value one for large positive $n$. At large $n\gg1$, the matter
Hamiltonian $\hat{H}_{\phi}(n)$ is trivially related to that of
standard quantum cosmology by (\ref{n}), and we will focus on the
gravitational part $\hat{H}$ in what follows.

\section{The $\gamma\to0$ Limit}

The fundamental difference between loop and standard quantum cosmology
is the space-time discreteness of the former framework, controlled by
the parameter $\gamma$ (not just by $\lP$ since this scale is present
and non-zero in both theories). If we formally take the limit
$\gamma\to0$, $n\to\infty$ in such a way that
$p(n,\gamma)=\frac{1}{6}\gamma \lP^2n$ according to (\ref{n}) is
fixed, arbitrary values of $p$ are allowed and the discrete
$p$-spectrum approaches a continuous one. In what follows,
$\gamma\to0$ will always be understood as this simultaneous limit.

We will now show that in this limit the constraint equation
(\ref{Hloop}) gives (\ref{HWdW}) if we identify the wave functions
$\psi(p,\phi)=s_{n(p)}(\phi)$ at any value of $n$ for a fixed
$\gamma\not=0$ (but only large values $n\gg1$ become important in the
limit). First, we note that the coefficients $k_n^{\pm}$ approach one
for large $n$ and so can be dropped. Next, we introduce a new discrete
function
\[
 t_n(\phi):=\gamma^{-1}\lP^{-2}\left(V_{\frac{1}{2}n}-
 V_{\frac{1}{2}n-1}\right) s_n(\phi)
\]
which in the above limit with the identification of $s_n(\phi)$ and
$\psi(p,\phi)$ reduces to $\frac{1}{2} \sqrt{p} \psi(p,\phi)=:
\tilde{\psi}(p,\phi)$. The new wave function $t$ is subject to the
constraint equation
\[
 3\left(\case{1}{4}(1+\gamma^{-2})(t_{n+8}-2t_n+t_{n-8})-
 (t_{n+4}-2t_n+t_{n-4})\right)= -\gamma\lP^2
 \left(V_{\frac{1}{2}n}- V_{\frac{1}{2}n-1}\right)^{-1}
 \kappa\hat{H}_{\phi}(n) t_n\,.
\]
Now we use (\ref{n}) to identify $t_{n+k}$ with $\tilde{\psi}(p+\Delta
p(k))$ for any $k\in\Z$, $k\ll n$ and Taylor expand assuming $\gamma$ to
be small:
\[
 t_{n+k}=\tilde{\psi}(p+\case{1}{6}k\gamma\lP^2)= \tilde{\psi}(p)+
 \case{1}{6} k\gamma\lP^2\tilde{\psi}'(p)+ \case{1}{72}
 k^2\gamma^2\lP^4 \tilde{\psi}''(p)+ O(\gamma^3)\,.
\]
The gravitational part of the constraint can be written as
\begin{eqnarray*}
 && \case{1}{4}(1+\gamma^{-2})(t_{n+8}(\phi)-2t_n(\phi)+t_{n-8}(\phi))-
 ( t_{n+4}(\phi)-2t_n(\phi)+t_{n-4}(\phi))\\
 && = \case{1}{4} (1+\gamma^{-2})
 \cdot\case{16}{9} \gamma^2\lP^4 \frac{\md^2}{\md p^2}
 \tilde{\psi}(p,\phi)- \case{4}{9} \gamma^2\lP^4
 \frac{\md^2}{\md p^2} \tilde{\psi}(p,\phi) +O(\gamma^4)\\
 &&= \case{4}{9}\lP^4 \frac{\md^2}{\md p^2}
 \tilde{\psi}(p,\phi)+O(\gamma^4)
\end{eqnarray*}
which in the limit $\gamma\to0$ yields (\ref{HWdW}):
\[
  \case{4}{3}\lP^4 \frac{\md^2}{\md p^2}
  \tilde{\psi}(p,\phi)= \case{2}{3}\lP^4 \frac{\md^2}{\md
  p^2} (\sqrt{p}\psi(p,\phi))= -\kappa\hat{H}_{\phi}\psi(p,\phi)\,.
\]

We see that standard quantum cosmology (in its full range of all
positive values of $p$) is the $\gamma\to0$ limit of loop quantum
cosmology, which shows the role of $\gamma$ as a useful constant
controlling the space-time discreteness. Of course, the physical value
of $\gamma$ is non-zero \cite{ABCK:LoopEntro,IHEntro} implying that
standard quantum cosmology is valid only in certain regimes at large
volume where the discreteness does not matter (the above expansion of
$t_{n+k}$ can still be done if $p\gg\lP^2$ and $\psi$ is not wildly
varying at the Planck scale, irrespective of the value of
$\gamma$). In general, there will be $\gamma$-corrections to standard
quantum cosmology which can be derived from loop quantum
cosmology. Only at very small volumes, close to the classical
singularity, will it not be sufficient just to include correction
terms in an effective Hamiltonian. In this regime, where standard and
loop quantum cosmology differ drastically, a discrete formulation is
inevitable.

\section{The Semiclassical Limit}

One way to demonstrate the correct classical limit of a quantum theory
consists in introducing the Wigner function \cite{Wigner} (adapted
here to our notation for an isotropic cosmological model; see also
\cite{Halliwell2} for an appearance of the Wigner function in standard
quantum cosmology)
\[
 W(p,c)= \int\md u \overline{\psi(p-\case{1}{2}\lP^2 u)}
 \exp(-icu) \psi(p+\case{1}{2} \lP^2 u)
\]
which associates a distribution on the classical phase space to any
state $\psi$. It is a probility distribution (i.e., a non-negative
function) only in the classical limit $\hbar\to0$, in which case the
quantum evolution equation $\hat{H}\psi=0$ (written as a Hamiltonian
constraint for a system with internal time) reduces to the classical
Liouville equation $\{H,W\}=0$ (see, e.g., \cite{Karasev}). For a
generalization to a discrete configuration space, on which
$\frac{1}{2}u$ may not be defined for all allowed $u$, the form
\begin{equation}\label{Wigner}
 \tilde{W}(p,c)=\overline{\psi(p)}\int\md u \exp(-icu)
 \psi(p+\lP^2 u)
\end{equation}
will be more suitable. Its properties differ from those of $W$
only at orders of $\hbar$ or higher which are irrelevant for the
semiclassical limit.

Wave packets which correspond to a single particle state (rather than
a statistical ensemble) in the classical limit can be constructed by
using the WKB approximation for a state $\psi$ written in the form
$\psi(p)=C(p)\exp(\pm i\hbar^{-1} S(p))$ with
$S(p)=S_0(p)+O(\hbar^2)$, where $S_0$ will turn out to be the
classical action, and an $\hbar$-independent function $C$.  An
expansion of $\psi(p\pm \case{1}{2}\lP^2u)$ shows that a state $\psi$
in the WKB form has a Wigner function $W(p,c)=|C(p)|^2 \delta(c\mp \md
S/\md p)+O(\hbar)$ which is peaked about classical solutions given by
$c(p)=\md S/\md p$.

In our case of (\ref{HWdW}) we choose an ansatz
$\psi(p,\phi)=p^{-\frac{1}{2}} C(p) \exp(\pm i\hbar^{-1}S_0(p))
\xi(\phi)$ where $\xi$ is a ($p$-dependent) eigenstate of the matter
Hamiltonian: $\hat{H}_{\phi}(p)\xi= E(p)\xi$. This means that only the
gravitational part is treated semiclassically, whereas matter is in a
quantum state depending adiabatically on $p$. Then we can neglect the
$p$-dependence of $\xi$, and (\ref{HWdW}) implies
\[
 \case{2}{3}(\kappa S_0')^2-\kappa p^{-\frac{1}{2}} E(p)\mp \case{2}{3}
 i\kappa\lP^2 (S_0''+2C^{-1}S_0'C')+ O(\lP^4)=0\,,
\]
and so $S_0(p)$ fulfills the Hamilton--Jacobi equation for the
gravitational background with matter energy $E(p)$, and we have
$C(p)=S_0'(p)^{-\frac{1}{2}}= \sqrt{\case{1}{3}\gamma\kappa
c(p)^{-1}}$. Note that unlike a kinematical semiclassical state, a
dynamical semiclassical state cannot be peaked about one phase space
point. In particular in an internal time formulation, a peak in the
phase space function chosen as internal time cannot be allowed.

The WKB approximation is valid in regimes where $\hbar\log C(p)$ is not
strongly varying compared to $S_0(p)$, and thus the $O(\lP^2)$-term is in
fact a small correction to the Hamilton--Jacobi equation. Then,
\begin{equation}\label{WKB}
 2\left|\hbar(S_0')^{-1}\frac{\md}{\md p}\log C\right|= \left|\frac{\hbar
 S_0''}{(S_0')^2}\right|=\left|\frac{\md\lambda}{\md p}\right|\ll1
\end{equation}
and $\psi\sim\exp(\pm i\hbar^{-1} S_0(p))$ can locally be written as a
wave with constant wave length $\lambda=\hbar(S_0')^{-1}$ defined by
\[
 \hbar^{-1} S_0(p_0+\Delta p)= \hbar^{-1}S_0(p_0)+ \hbar^{-1}
 S_0'(p_0)\Delta p+ O(\Delta p^2)=: \hbar^{-1}S_0(p_0)+ \lambda^{-1}
 \Delta p+O(\Delta p^2)\,.
\]

Since $\psi$ is a wave function on (mini-)superspace and
$\lambda$ gives the oscillation length in $p$ (therefore, $\lambda$
has dimension length$^2$), which is discrete at a fundamental
level, we obtain an additional condition: the oscillation length
cannot be smaller than the smallest possible scale, which gives a new
condition
\begin{equation} \label{Scale}
 |\lambda|\gg\lP^2
\end{equation}
in addition to (\ref{WKB}). Otherwise, a state would have strong
variation between successive values of $n$ violating the
pre-classicality condition of \cite{DynIn}. (Note that this condition
may even be violated at large volume, e.g.\ in the presence of a
cosmological constant $\Lambda$ in which case we have an action
$S_0\propto \sqrt{\Lambda}V=\sqrt{\Lambda p}\cdot p$ leading to a wave
length which tends to zero for large $p$. This kind of infrared
problem, however, is an artefact of the minisuperspace approximation
--- curvature enters the formalism in the space integrated form which
can be large even if the local curvature scale is small --- and can be
ignored here.)

When the two conditions (\ref{WKB}), (\ref{Scale}) for a semiclassical
behavior are fulfilled, the correct semiclassical limit of loop
quantum cosmology follows from the discussion above: the continuum
limit $\gamma\to0$ leads to standard quantum cosmology whose
semiclassical limit $\hbar\to0$ is demonstrated using the Wigner
function. It is also possible to sidestep the Wheeler--DeWitt
formulation by defining a Wigner function directly for loop quantum
cosmology. A straighforward generalization of (\ref{Wigner}) is
\[
 W_n(c)=(\sqrt{2}\sin(\case{1}{2}c))^{-1} s_n \sum_{k\in\Z}
 T_k(c)s_{n+k}
\]
which is a distribution function, associated with a state $s_n$, on
the (partially discrete) space with coordinates $(n,c)$. Because of
the discreteness, the integral in (\ref{Wigner}) is replaced by a
sum. Moreover, instead of the Fourier transform we use the functions
$T_n(c)= (\sqrt{2}\sin(\case{1}{2}c))^{-1} \exp(\case{1}{2}inc)$ which
are eigenstates of $\hat{p}$ and are used in a transformation between
the triad and the connection representation \cite{IsoCosmo}. The
factor $(\sqrt{2}\sin(\case{1}{2}c))^{-1}$ in $W_n(c)$ accounts for
the non-trivial measure in the $c$-representation. The classical limit
again is derived by an expansion of the constraint equation for
$W_n(c)$ in both $\gamma$ and $\hbar$. Note that $\gamma$ always
appears multiplied by $\hbar$ (or $\lP^2$), and so the discreteness
corrections also disappear if we perform the limit $\hbar\to0$ by
itself, fixing $\gamma$. However, if we are interested not only in the
classical limit but also in corrections, two different types of terms
occur: one caused by the discreteness (depending on $\gamma$) and one
purely from quantum theory ($\gamma$-independent). Unlike before, the
continuum limit $\gamma\to0$ is now done at the phase space level and
standard quantum comology does not appear as an intermediate step.

We can directly apply the WKB prescription to the discrete wave
function $s_n\sim C(n)\exp(\pm i\hbar^{-1}S(p(n)))$. Because we
already know that the constraint equations of standard and loop
quantum cosmology agree up to $\gamma$-corrections, this state is an
approximate solution to the constraint (\ref{Hloop}) up to $\hbar$-
{\em and\/} $\gamma$-corrections. Locally (in a neighborhood of a
given value $n_0$), for any solution $s_n$ of (\ref{Hloop}) there is a
solution $\psi(p)$ of (\ref{HWdW}) which approximates $s_n$. But the
$\gamma$-corrections in general add up when solving the difference
equation, and so two solutions $s_n$ and $\psi(p)$ can differ, e.g.\
by a phase shift, away from $n_0$. Nevertheless, the leading order (in
$\gamma$ and $\hbar$) of the Wigner functions associated with both
states is the same, which is a direct consequence of the expansion
being sensitive only to the local behavior of a wave function.

\section{Conclusions}

We have shown that one can perform the semiclassical limit of
isotropic, spatially flat loop quantum cosmology in a two-step
procedure leading to the correct classical behavior: the first step is
the continuum limit which can be formulated as $\gamma\to0$, followed
by a second step which is the usual semiclassical limit of quantum
mechanics. At the intermediate level, one obtains standard quantum
cosmology as the continuum limit of loop quantum cosmology. This also
shows how to define semiclassical (WKB) states at the dynamical level
of isotropic quantum cosmology whose correlations are peaked about the
classical ones.

We conclude with a few remarks about lessons for a possible
generalization to the full theory of quantum geometry. In this case
the inhomogeneity is manifested in the appearance of arbitrary graphs
to which states are associated. The continuum limit can then no longer
be performed by only $\gamma\to0$ together with labels going to
infinity, but has to be extended by a prescription of how to shrink
the graphs to continuous objects (analogous to a vanishing lattice
spacing in lattice gauge theories). In this process the number of
vertices will become infinite, and the standard methods of quantum
geometry become ill-defined; this may be related to the fact that the
Wheeler--DeWitt quantization of gravity is only defined formally for
inhomogeneous geometries. In this respect, the second strategy of the
previous section, defining a Wigner function directly on the discrete
phase space, will be more suitable since it does not make use of an
intermediate Wheeler--DeWitt quantization.

In addition to the issue of graphs, the full Hamiltonian constraint
operator is more complicated than that of isotropic models. This is in
part due to the appearance of arbitrary graphs, but most importantly
due to a complicated volume spectrum which is not known explicitly. A
possible route consists in again using reduced models, this time
midi-superspace models which are inhomogeneous (e.g., spherically
symmetric models or cylindrical waves which both have a single
inhomogeneous axis). This allows to investigate the new aspects of
inhomogeneous states without dealing directly with the full
constraint.

Some observations can already be inferred from the classical form of
the Hamiltonian constraint with Euclidean part
\begin{equation} \label{Eucl}
 \kappa\det(e_I^i)\HE=-\epsilon_{ijk}F_{ab}^iE_j^aE_k^b=
 -2\left(\epsilon_{ijk} \partial_{[a}A_{b]}^i+ A_{[a}^jA_{b]}^k\right) 
 E^a_jE^b_k
\end{equation}
and extrinsic curvature part
\begin{equation} \label{Ext}
 \kappa\det(e_I^i)P=-2(1+\gamma^{-2})(A_a^i-\Gamma_a^i)
 (A_b^j-\Gamma_b^j) E_i^{[a}E_j^{b]}
\end{equation}
which yield the constraint
\begin{eqnarray*}
 H &=& -\HE+P\\
 &=& 2\kappa^{-1}\det(e_I^i)^{-1}\left(\epsilon_{ijk} \partial_a
 A_b^k- \gamma^{-2} A_{[a}^iA_{b]}^j- (1+\gamma^{-2}) \left(\Gamma^i_{[a}
 \Gamma^j_{b]}-2A_{[a}^i\Gamma_{b]}^j\right)\right) E^a_iE^b_j\,.
\end{eqnarray*}
In the isotropic, spatially flat constraint only the middle term is
present which leads to a $\gamma$-independent constraint after
quantizing in a triad representation since $A_a^i$ becomes an operator
proportional to $\gamma$. The first term, on the other hand, would be
$\gamma$-dependent, but also is sensitive to the continuum limit of
graphs (due to the partial derivative). This suggests to link the
graph continuum limit to the $\gamma\to0$ limit. The last term
containing $\Gamma_a^i$ looks problematic since it diverges in the
$\gamma\to0$ limit. However, locally one can always choose coordinates
such that $\Gamma_a^i=0$ eliminating this term. Since invariance under
coordinate transformations is incorporated by the diffeomorphism
constraint, this term may play a role in understanding the relation
and algebra of the quantized diffeomorphism and Hamiltonian
constraints.

As another application one can use similar ideas to find a relation
between the loop and Fock space quantizations of Maxwell theory, which
also allows an analog of the Barbero--Immirzi parameter (usually set
to one by using the known value of the fundamental charge
\cite{MaxwellSpinnet}). This would lead to a transformation from a
quantization with discrete charges to one with a continuous charge
spectrum, different from \cite{Fock}.

\section*{Acknowledgements}

The author is grateful to A.\ Ashtekar and H.\ Morales-T\'ecotl for
discussions.  This work was supported in part by NSF grant PHY00-90091
and the Eberly research funds of Penn State.

\end{document}